\documentclass[conference]{IEEEtran}
\IEEEoverridecommandlockouts
\usepackage{cite}
\usepackage{amsthm}
\usepackage{amsmath,amssymb,amsfonts}
\usepackage{algorithmic}
\usepackage{graphicx}
\usepackage{textcomp}
\usepackage{xcolor}
\usepackage{booktabs}
\usepackage{tabularx}
\usepackage{graphicx}
\usepackage{multicol}

\newtheorem{proposition}{Proposition}[section]
\newtheorem{theorem}{Theorem}[section]

\def\BibTeX{{\rm B\kern-.05em{\sc i\kern-.025em b}\kern-.08em
    T\kern-.1667em\lower.7ex\hbox{E}\kern-.125emX}}

\begin{document}

\title{UAV-enabled Computing Power Networks: Task Completion Probability Analysis\\ 
}


\author{\IEEEauthorblockN{Yiqin Deng, Zhengru Fang, Senkang Hu, Yanan Ma, Haixia Zhang, and Yuguang Fang,~\IEEEmembership{Fellow,~IEEE}
}
}

\maketitle

\begin{abstract}
This paper presents an innovative framework that synergistically enhances computing performance through~\textit{ubiquitous computing power distribution} and~\textit{dynamic computing node accessibility control} via adaptive unmanned aerial vehicle (UAV) positioning, establishing~\textit{UAV-enabled Computing Power Networks (UAV-CPNs)}.
In UAV-CPNs, UAVs function as dynamic aerial relays, outsourcing tasks generated in the~\textit{request zone} to an expanded~\textit{service zone}, consisting of a diverse range of computing devices, from vehicles with onboard computational capabilities and edge servers to dedicated computing nodes. This approach has the potential to alleviate communication bottlenecks in traditional computing power networks and overcome the ``island effect'' observed in multi-access edge computing.
However, how to quantify the network performance under the complex spatio-temporal dynamics of both communication and computing power is a significant challenge, which introduces intricacies beyond those found in conventional networks. To address this, in this paper, we introduce \textit{task completion probability} as the primary performance metric for evaluating the ability of UAV-CPNs to complete ground users' tasks within specified end-to-end latency requirements. Utilizing theories from stochastic processes and stochastic geometry, we derive analytical expressions that facilitate the assessment of this metric. Our numerical results emphasize that striking a delicate balance between communication and computational capabilities is essential for enhancing the performance of UAV-CPNs. Moreover, our findings show significant performance gains from the widespread distribution of computing nodes.
\end{abstract}


\section{Introduction}
The proliferation of latency-sensitive and computation-intensive applications, such as augmented reality and autonomous driving~\cite{Fang2025, Hu2025}, has intensified the demand for ubiquitous networks that seamlessly integrate communication and distributed computing capabilities. While edge computing architectures have expanded computational capacity at the network edge, existing computing resources remain geographically fragmented, leading to resource under-utilization and creating computing power islands, known as ``island effect''~\cite{Sun2024}. The International Telecommunication Union (ITU) has recently introduced the novel concept of computing power networks to address these challenges. The aim is to build a computation-driven network based on existing communication infrastructures. However, ensuring the stability and efficiency of computing power networks faces many design challenges. One major issue is the communication bottleneck in providing ubiquitous coverage with ground users (GUs) on a wide range of mobile devices and in accessing isolated computing power islands (e.g., edge servers, computing clusters, and user-provided computing nodes). Another challenge is the cost associated with the ubiquitous deployment of computing nodes (CNs)~\cite{Sun2024, ma2025uav}. How to build a cost-effective connected computing power network to reach geographically distributed computing power resources is a challenging but important design task.  

The advancement of low-altitude economy initiatives offers opportunities to leverage existing aerial facilities, such as unmanned aerial vehicles (UAVs), to carry communication or computing equipment, thereby providing flexible and low-cost services. In this context, we propose~\textit{UAV-enabled Computing Power Networks (UAV-CPNs)}, where UAVs serve as dynamic aerial relays to connect spatially distributed and heterogeneous computing resources into a unified service architecture. This enables UAVs to relay heavy computing tasks to remote terrestrial CNs. 
This framework is particularly effective in urban core areas, where traditional infrastructure often struggles with peak-hour communication and computing congestion due to dense GU demands or severe path loss from non-line-of-sight (NLoS) propagation through urban canyons. A complete task execution cycle in UAV-CPNs comprises three sequential phases:
\begin{enumerate}
    \item \textit{GU-to-UAV task offloading}: GUs transmit task data to the UAV.
    \item \textit{UAV-to-CN forwarding}: The UAV forwards tasks to a CN for computing.
    \item \textit{CN computing and result return}: The CN computes the tasks and returns the results to GUs.
\end{enumerate}

Existing studies on UAV-assisted multi-access edge computing/ computing power networks~\cite{Dong2024, tao2024multi, ma2025uav} typically relies on static CN accessibility models, where the number of available CNs remains fixed regardless of UAV operational positioning. This rigid assumption fundamentally constrains performance improvements when extending CNs for computing beyond geographically constrained demand areas, as it fails to use potentially abundant but spatially distributed computing resources outside the task request areas~\cite{deng2024uav}. Unlike these conventional approaches, the proposed UAV-CPN architecture exhibits two defining characteristics. First, computing power is ubiquitously available across larger geographical areas through heterogeneous CNs, ranging from vehicles (e.g., connected and autonmous vehicles or CAVs) with onboard computational capabilities~\cite{Chen2024} to dedicated CNs. 
Second, strategic UAV positioning (altitude and horizontal coordinates) dynamically regulates CN accessibility, a concept explored in our prior work~\cite{Deng2024}, where we assume that all CNs are located within the task request area.
To illustrate the altitude-dependent trade-offs, consider vertical UAV positioning. Higher altitudes improve line of sight (LoS) probability between GUs and UAVs, improving uplink reliability for \textit{GU-to-UAV task offloading} and downlink reachability to remote CNs. However, increased path loss and propagation delay due to longer transmission distances may degrade signal quality and increase communication latency. This creates a non-convex optimization landscape for UAV altitude selection. For example, the altitude that maximizes uplink transmission rates for \textit{GU-to-UAV task offloading} may result in suboptimal performance for the subsequent \textit{UAV-to-CN forwarding} and \textit{CN computing}. This is because downlink transmission rates depend on the quality of the UAV-CN communication channel, while the accessibility of CN is constrained by the UAV’s coverage area, both of which are also influenced by the UAV’s  altitude. 
When considering horizontal UAV positioning, these interdependencies evolve into multi-dimensional spatio-temporal trade-offs. For instance, positioning closer to high-density GU areas reduces uplink delays but may limit access to high-capacity CNs beyond GU coverage. Current research inadequately addresses these compounded effects.

In this paper, we propose a novel framework for UAV-CPNs that diverges from traditional air-ground integrated networks by enabling an unbounded number of CNs to participate in task processing, provided that the latency requirements are met. Specifically, we present an innovative network spatial model that simultaneously captures the dynamics of UAVs, the distribution of GUs, and CNs. By employing a probabilistic air-ground channel model, we characterize the dynamic communication conditions in terms of UAV altitude for both \textit{GU-to-UAV task offloading} and \textit{UAV-to-CN forwarding}. Moreover, we characterize the computing latency dynamics on CNs through a probabilistic model,  yielding a generalized framework for heterogeneous computing environments. 

Focusing on a foundational scenario with a single UAV and its vertical positioning (due to scope constraints), we define \textit{task completion probability} as the critical performance metric, quantifying the system's ability to complete GUs' tasks under E2E latency constraints. By integrating stochastic process theory with tools from stochastic geometry, we derive analytical expressions for this probability. Numerical simulations reveal trade-offs between communication parameters (e.g., UAV altitude) and computing parameters (e.g., CN density and coverage), exposing potential bottlenecks in resource allocation. These findings establish a theoretical foundation for the seamless convergence of communication and computing in future-generation networks. 

\section{Modeling UAV-enabled computing power networks}
\subsection{Network Spatial Model}

\begin{figure}[t]
    \centering
    \includegraphics[width=0.8\linewidth]{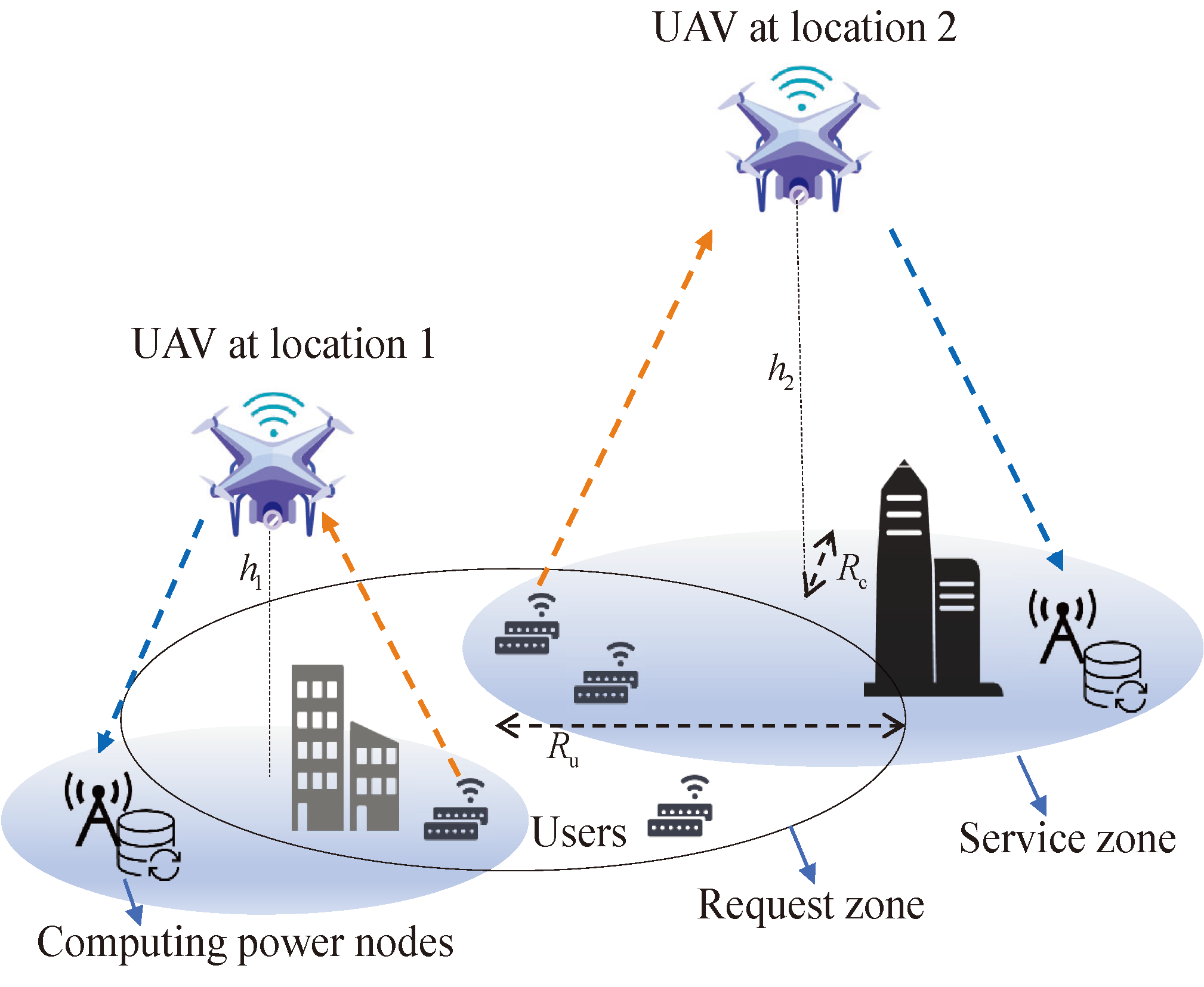}
    \caption{Illustration of a UAV-enabled computing power network, where GUs offload tasks generated within the service area to distributed computing power nodes for processing. The computing accessibility is enhanced by strategically adjusting key network parameters, e.g., the position of the aerial UAV relay. }
    \label{fig:system}
    \vspace{-10pt} 
\end{figure}

As illustrated in Fig.~\ref{fig:system}, we consider a circular area with a fixed radius $R_u$, referred to as the \textit{request zone}, an area of interest for task request generations, where GUs follow a uniform spatial distribution. CNs are distributed over an unbounded plane, encompassing the \textit{request zone}, according to a homogeneous Poisson Point Process (PPP) $\Phi_C$ with density $\lambda_c$ (nodes/m$^2$). The GU and CN spatial distributions are mutually independent, i.e., task generations and CN node availability are mutually independent.

UAVs operate at altitude $h$ above the \textit{request zone}, serving as aerial relays for two-way computation offloading from GUs to CNs. The combined coverage area of CNs and UAVs defines the \textit{service zone}, an area where CNs can be found to serve tasks, whose effective size depends on the UAV deployment parameters. While the \textit{service zone} is notionally unbounded due to the infinite PPP support, practical accessibility is constrained by UAV positioning and quality of communication channels from UAV to GUs and CNs. The expected number of GUs or CNs in any subregion equals the product of their respective densities and the subregion's area due to the PPP assumption. 

The maximum~\textit{service zone} radius is constrained by the worst-case scenario where the cumulative transmission latency across~\textit{GU-to-UAV task offloading} and~\textit{UAV-to-CN task forwarding} reaches the total allowable latency budget $T_{\text{max}}$, thereby ensuring strict compliance with E2E quality-of-service (QoS) requirements.

\subsection{Air-ground Channel Model}
Following~\cite{Mozaffari2016}, we adopt a probabilistic air-ground channel model to characterize line-of-sight (LoS) and non-line-of-sight (NLoS) connectivity probabilities for both~\textit{GU-to-UAV task offloading} and~\textit{UAV-to-CN task forwarding} links. 

In the \textit{GU-to-UAV task offloading} phase, the transmit power of GUs is $ P_u $, the received power at the UAV is expressed as:
\begin{equation}\label{eq:up_power}
     P_{r,\text{up}} = \begin{cases} 
     P_u \cdot \left(r_u^2 + h^2\right)^{-\alpha_u/2} & \text{(LoS)}, \\
     \eta P_u \cdot \left(r_u^2 + h^2\right)^{-\alpha_u/2} & \text{(NLoS)},
     \end{cases}
\end{equation}
where $ r_u $ is the horizontal GU-UAV distance, $ \alpha_u $ is the uplink path loss exponent, and $ \eta \in (0,1) $ is the NLoS attenuation factor. The probability of LoS connectivity is given by:
\begin{equation}
     P_{\text{LoS,up}} = \frac{1}{1 + C \exp\left(-B\left(\frac{180}{\pi}\arctan\left(\frac{h}{r_u}\right) - C\right)\right)},
\end{equation}
where $ B $ and $ C $ are environment-dependent parameters. 

Analogously, for~\textit{UAV-to-CN task forwarding}, the received power at a CN, denoted as \( P_{r, \text{down}} \), can be modeled based on  the transmit power of the UAV \( P_d \), the horizontal UAV-CN distance \( r_c \), and the downlink path loss exponent \( \alpha_c \).


\subsection{Computing Model}
We consider heterogeneous CNs cooperatively operated by multiple service providers, where some factors such as hardware heterogeneity, queuing delays, and I/O interference among virtual machines collectively influence computational throughput~\cite{Huang2018}. To generically model the dynamic computing capabilities of CNs, reflecting their uncertainty as well, we characterize the computation time \( t_c \) through its cumulative distribution function (CDF):
\begin{equation}\label{eq:comp_latency}
F_{t_c}(T_{\text{res}}; D) = \mathbb{P}(t_c \leq T_{\text{res}}), 
\end{equation}
where  \( T_{\text{res}} = T_{\text{max}} - T_{\text{comm}} \) is the residual time budget after accounting for total communication latency \( T_{\text{comm}} \),  and \( D \) denotes the task-specific computational workload. This CDF quantifies the probability that a CN completes the task within \( T_{\text{res}} \) with workload \( D \). 

\section{Task completion probability analysis}
In this section, we focus on a scenario where a UAV deployed at altitude \( h \) above the centroid of a circular \textit{request zone} with radius \( R_u \) forwards a single GU's computational tasks (with data size \( D \)) to a CN within the \textit{service zone}.
This single-CN assumption establishes a theoretical performance floor for the system, providing a conservative performance benchmark that intentionally excludes performance gains from multi-CN parallel processing, a common optimization strategy in existing literature~\cite{Deng2024}. To investigate the fundamental communication-computing tradeoff, we adopt idealized conditions without consideration of multi-user interference and resource contention, leaving such analyses for future extensions due to page limit.  

We begin with modeling the E2E latency for task processing in the UAV-CPN framework. Building on this model, we formally define the task completion probability and derive its analytical expression for a GU at an arbitrary location within the \textit{request zone}. We then generalize this metric by spatially averaging over all GU positions to characterize system-wide performance. Through these analyses, we reveal the fundamental trade-offs among several critical parameters such as the CN density \( \lambda_c \), the UAV's operational altitude \( h \), and the latency budget \( T_{\text{max}} \).
The results demonstrate how these parameters jointly determine the operational success of UAV-CPN task computing, yielding actionable design insights for latency-sensitive deployments.

\subsection{End-to-end latency}
Given the bandwidth $W$ and noise power $N_0$, the transmission latency from the GU to the UAV ($t_1$) can be calculated by:
\begin{equation}
t_1 = \frac{D}{W \log_2 \left(1 + {P_{r,\text{up}}}/{N_0}\right)}.
\end{equation}
Similarly, the transmission latency from the UAV to the CN ($ t_2 $) can be calculated using $P_{r,\text{down}}$.
The E2E latency represents the total time from task generation at the GU to result reception. As assumed in most prior works, we ignore result feedback latency due to the small size of the result, yielding the total E2E latency as $T_{\mathrm{E2E} }=t_1+t_2+t_c$. 

\subsection{Performance metric}
The task completion probability serves as a critical performance metric, reflecting both the communication and computational capabilities of the system, which can be used to compute other performance metrics. For a specific GU, we define the task completion probability as the likelihood of locating a CN within the coverage area of a UAV to complete the computing task at this CN within the E2E latency constraint. Specifically, this means satisfying the condition $ t_1 + t_2 + t_c \leq T_{\text{max}} $. For system-wide analysis, this metric is spatially averaged over all GU and CN positions governed by their respective distributions. 

If \( t_1 + t_2 \geq T_{\text{max}} \), the accumulated transmission latency across both the \textit{GU-to-UAV task offloading} and \textit{UAV-to-CN task forwarding} phases exceeds the latency budget, which results in a communication bottleneck, termed the \textit{comm-limited} scenario. Instead, if \( t_c \geq T_{\text{max}} - t_1 - t_2 \), the computational latency at the CN exceeds the residual time budget \( T_{\text{res}} \), resulting in a computational bottleneck termed the \textit{comp-limited} scenario.  
Task completion probability necessitates the concurrent fulfillment of both communication and computational constraints. Violation of either constraint, whether \textit{comm-limited} or \textit{comp-limited}, severely degrades system performance. Leveraging stochastic geometry, we derive analytical expressions for the task completion probability.  

\subsection{Bottleneck Analysis}
For a GU at horizontal distance \( r_u \) from the UAV, the latency constraint \( T_{\text{max}} \) fundamentally limits service accessibility despite the theoretical availability of all CNs in UAV-CPNs. Specifically, the \textit{comm-limited} condition imposes a critical spatial restriction by bounding the maximum   radius \( r_c^{\mathrm{max}}(r_u) \) for the effective \textit{service zone}, which is determined by:
\begin{equation}\label{eq:max_coverage}
t_2=t_2\left(r_c^{\mathrm{max}}(r_u)\right) = T_{\text{max}} - t_1(r_u),
\end{equation}
where $t_1=t_1(r_u)$ and $t_2=t_2(r_c)$ are the transmission times for~\textit{GU-to-UAV task offloading} and~\textit{UAV-to-CN task forwarding}, respectively.

Within this communication-constrained~\textit{service zone} (\( r_c \leq r_c^{\mathrm{max}}(r_u) \)), CNs must additionally satisfy the computing latency requirement. For residual time budget \( T_{\text{res}} \triangleq T_{\text{max}} - t_1 - t_2 \), the probability of a CN satisfying this latency constraint is quantified through its CDF as \( F_{t_c}(T_{\text{res}}; D) \), as characterized by Eq.~\eqref{eq:comp_latency}.  


Based on previous analysis, we establish that task completion fails under either \textit{comm-limited} or \textit{comp-limited} conditions. Specifically, transmission failure occurs when CNs reside outside the communication-constrained~\textit{service zone} (\( r_c \leq r_c^{\mathrm{max}}(r_u) \)), leading to failure in task processing during the transmission phase. For successfully transmitted tasks, the task completion probability equals the likelihood of satisfying the computing latency requirement within the residual time budget $T_{\text{res}}$. These dual constraints induce a \textit{probabilistic thinning}~\cite{Lalley20XX} of the original homogeneous PPP \( \Phi_c \), producing a thinned PPP with spatially varying density. The resultant effective CN density is formally characterized in Proposition~\ref{prop:density}.

\begin{proposition}[Effective CN Density]
\label{prop:density}
For a GU at a horizontal distance \( r_u \) from the UAV, the spatial density of CNs satisfying E2E latency constraint, simply called effective density or conditional spatial density, is given by:
\begin{equation}
\lambda_c^{\mathrm {eff}}(r_u) =\lambda_c  \mathbb{I}_{\{r_c \leq r_c^{\mathrm{max}}(r_u)\}} \cdot F_{t_c}(T_{\text{res}}; D),
\end{equation}
where \( \mathbb{I}_{\{r_c \leq r_c^{\mathrm{max}}(r_u)\}} \) indicates successful communication; \( F_{t_c}(T_{\text{res}}; D) \), as the CDF of computing latency, quantifies the probability of computing latency within $T_{\text{res}}$.
\begin{proof}
A qualified CN must be located within the maximum coverage radius \( r_c^{\mathrm{max}}(r_u) \) for the \textit{service zone} and have a computing latency \( t_c \leq T_{\text{res}} \). The former condition is modeled by an indicator function, referred to as \textit{success in communication}. The latter models the probability that a CN satisfies the computing latency constraint, referred to as \textit{success in computing}. Since the spatial distribution and computing power dynamics are independent, the density of qualified CNs for task completion is the product of these two probabilities. Thus, each CN at distance $r_c$ is retained with probability:
\begin{equation}
p_{\text{retain}}(r_u) = 
\underbrace{\mathbb{I}_{\{r_c \leq r_c^{\mathrm{max}}(r_u)\}}}_{\text{Success in communication}} 
\cdot 
\underbrace{ F_{t_c}(T_{\text{res}}; D)}_{\text{Success in computing}}.
\end{equation}
This qualification process represents an \textit{independent thinning} of the original PPP $\Phi_c$. By the Poisson thinning property~\cite{Lalley20XX}, the resulting spatially thinned process retains the PPP property with effective density:
\begin{equation}
\lambda_c^{\mathrm{eff}}(r_u) = \lambda_c \cdot p_{\text{retain}}(r_c).
\end{equation}
\end{proof}
\end{proposition}
Since the thinning process preserves the PPP properties, the qualified CNs form a thinned PPP with effective density \(\lambda_c^{\mathrm{eff}}(r_u)\). The expected number of qualified CNs is derived by integrating this effective density over the spatial domain of interest, as formalized in Proposition~\ref{prop:number}. 

\begin{proposition}[Qualified Numbers of CNs]
\label{prop:number}
The expected number of qualified CNs for a GU located at a horizontal distance $ r_u $ from the UAV is given by:
\begin{equation}\label{eq:intensity}
\Lambda(r_u) = 2\pi\lambda_c \int_0^{r_c^{\mathrm{max}}(r_u)}  F_{t_c}(T_{\text{res}}; D)r_c \, dr_c.
\end{equation}
\begin{proof}
The expected number of qualified CNs is derived from the intensity measure of the thinned PPP:
\begin{equation}\label{eq:integral}
\Lambda = \iint_{\mathbb{R}^2} \lambda_c^{\mathrm{eff}}(\|\boldsymbol{x}\|) d\boldsymbol{x}.
\end{equation}
Exploiting circular symmetry about the UAV, we convert to polar coordinates ($d\boldsymbol{x} = r_c dr_c d\theta$):
\begin{equation}
\begin{aligned}
\Lambda(r_u) &= \int_0^{2\pi} \int_0^{r_c^{\mathrm{max}}(r_u)} \lambda_c^{\mathrm{eff}}(r_c) r_c \, dr_c \, d\theta \\
&= 2\pi \lambda_c \int_0^{r_c^{\mathrm{max}}(r_u)}  F_{t_c}(T_{\text{res}}; D) r_c \, dr_c.
\end{aligned}
\end{equation}\qedhere
\end{proof}
\end{proposition}

\subsection{Task completion probability}
Based on the previous analysis, the task completion probability for a specific GU is summarized in  Theorem~\ref{theorem:success}.
\begin{theorem}\label{theorem:success}
For a GU located at a horizontal distance $ r_u $ from the UAV, the probability that it can associate with at least one CN satisfying the E2E latency constraint $ t_1 + t_2 + t_c \leq T_{\text{max}} $ is given by:
\begin{multline}\label{eq:success}
P_{\text{success}}(r_u) = 1 - \exp\biggl( -2\pi \lambda_c \\
\times \int_0^{r_c^{\mathrm{max}}(r_u)}  F_{t_c}\left(T_{\text{max}} - t_1(r_u) - t_2(r_c); D\right) r_c \, \mathrm{d}r_c \biggr),
\end{multline}
where $ t_1(r_u) $, $ t_2(r_c) $, $ r_c^{\mathrm{max}}(r_u) $, and $ F_{t_c}\left(T_{\text{max}} - t_1(r_u) - t_2(r_c); D\right)$ are obtained from Eqs.~\eqref{eq:up_power}-\eqref{eq:max_coverage} when we consider the GU's location, i.e., the location at a horizontal distance $ r_u $ from the UAV.

\begin{proof}
From Proposition~\ref{prop:number}, the number of qualified CNs, for a GU located at a horizontal distance $ r_u $ from the UAV, follows a Poisson distribution with mean \( \Lambda(r_u) \). The void probability (i.e., no qualified CNs exist) is \( \exp(-\Lambda(r_u)) \)~\cite{Møller_Schoenberg_2010}. Thus, the success probability, which is the probability that there exists at least one CN that can complete the task, becomes: $P_{\text{success}}(r_u) = 1 - \exp(-\Lambda(r_u))$. Substituting $\Lambda(r_u)$ from Eq.~\eqref{eq:intensity} completes the proof. \qedhere
\end{proof}
\end{theorem}

\textbf{Remark:} \textit{(GU Location Dependency and CN Density Impact.)} From Theorem~\ref{theorem:success}, we observe that for a given UAV altitude \( h \), the success probability \( P_{\text{success}}(r_u) \) decreases as the GU's distance \( r_u \) from the UAV increases. This decrease occurs because GUs farther from the UAV experience longer offloading delays (\( t_1 \propto \log(r_u^2 + h^2) \)), which reduces the residual time budget available for forwarding (\( t_2 \)) and computing (\( t_c \)).
Moreover, for a given GU location, a higher CN density \( \lambda_c \) increases the spatial availability of CNs, thereby improving the task completion probability. Specifically, the task completion probability improves exponentially with the void probability of the PPP, given by \( \exp(-\Lambda(r_u)) \).

To evaluate system-level performance, we must consider the spatial distribution of GUs. The overall task completion probability, accounting for these factors, is provided in the following theorem (Theorem~\ref{theorem:average_probability}).

\begin{theorem}\label{theorem:average_probability}
The spatially averaged task completion probability for GUs uniformly distributed within the~\textit{request zone} (radius \( R_u \)) is: 
\begin{equation}
\overline{P}_{\text{success}} = \frac{2}{R_u^2}\int_0^{R_u} P_{\text{success}}(r_u)\cdot r_u \, dr_u,
\end{equation}
where \( P_{\text{success}}(r_u) \) is defined in Eq.~\eqref{eq:success}.   
\begin{proof} 
Owing to the uniform spatial distribution of GUs, the radial distance \( r_u \) follows the probability density function (PDF):  
\begin{equation}  
f_{r_u}(r_u) = \frac{2r_u}{R_u^2}, \quad 0 \leq r_u \leq R_u.  
\end{equation}  
By the law of total probability, the system-wide task completion probability equals the expectation of \( P_{\text{success}}(r_u) \) over this distribution:  
\begin{equation}  
\overline{P}_{\text{success}} = \mathbb{E}_{r_u}\left[ P_{\text{success}}(r_u) \right] = \int_0^{R_u} P_{\text{success}}(r_u) f_{r_u}(r_u) \, dr_u.  
\end{equation}  
Substituting \( f_{r_u}(r_u) \) completes the derivation.    
\quad \qedhere
\end{proof}
\end{theorem}

\textbf{Remark:} \textit{(Analytical Intractability.)} The nested integrals in $\overline{P}_{\text{success}}$ cause the difficulty in obtaining closed-form solutions due to three fundamental challenges. First, the coverage radius $r_c^{\mathrm{max}}(r_u)$ of the communication-effective~\textit{service zone} exhibits implicit nonlinear dependence on $r_u$ via the latency constraint (Eq. \eqref{eq:max_coverage}), coupling the integration domains of GU and CN locations. This interdependency prevents decoupling into separable integrals over $r_u$ and $r_c$. Second, the residual time budget $T_{\text{res}}=T_{\text{max}} - t_1(r_u) - t_2(r_c)$ within the integrand $F_{t_c}(T_{\text{res}}; D)$ inherits complexity from both components: $t_1(r_u)$ contains a logarithmic term $\log(r_u^2 + h^2)$, while $t_2(r_c)$, though ostensibly a function of $r_c$, indirectly depends on $r_u$ through the latency-constrained integration upper limit $r_c^{\mathrm{max}}(r_u)$. Third, the computing latency CDF $F_{t_c}(T_{\text{res}}; D)$ itself could be non-trivial. For instance, if $t_c$ models queueing delays in stochastic computing systems, its CDF may lack explicit analytical forms. These intertwined nonlinearities in spatial, temporal, and statistical dimensions lead us to resort to numerical integration techniques or stochastic geometry-based approximations for practical evaluation.

\section{Numerical results}
In this section, we evaluate the task completion probability using parameters listed in Table~\ref{table:parameter} unless explicitly indicated. These parameters align with the scenarios where UAVs and GUs employ wireless transmission modules to support low-latency services like real-time video analytics~\cite{Fang2025}. Due to the page limit, our evaluation focuses on three critical aspects: i) the validation of theoretical models, ii) the trade-off of communication-computing resources through strategic positioning of UAV altitude, and iii) the potential for performance enhancement with more reachable CNs. 

\begin{table}[ht]\label{table:parameter}
    \centering
    \setlength{\abovecaptionskip}{-2.5pt} 
    \setlength{\belowcaptionskip}{-5pt} 
    \caption{Parameter setting}
    \label{tab:simulation_parameters}
    \resizebox{\linewidth}{!}{
    \begin{tabular}{ccc}
        \toprule
        \textbf{Description} & \textbf{Parameter} & \textbf{Value} \\
        \midrule
        Transmit power of GUs and UAV &  $P_u$, $P_d$ & $20$ dBW, $20$ dBW \\
        Path loss exponent & $\alpha_u$ & $2$ \\
        NLoS attenuation coefficient & $\eta$ & $20$dB \\
        Bandwidth & $W$ & $8$ MHz \\
        Noise power & $N_0$ & $-120$ dBm \\
        Data size & $D$ & $1$ MB \\
        Maximum allowed latency & $T_{\text{max}}$ & $55$ ms \\
        GU density & $\lambda_u$ & $500$ nodes/km$^2$ \\
        CN density &$\lambda_c$ & $5$ nodes/km$^2$ \\
        Computing latency &$t_c$& $0.2$ ms\\
        Request zone radius & $R_u$& $200$m \\
        Parameters for urban environment & $B$, $C$ & $0.136$, $11.95$\\
        \bottomrule
    \end{tabular}}
\end{table}

\begin{figure}[b]
    \vspace{-18pt} 
    \centering
    \includegraphics[width=0.8\linewidth]{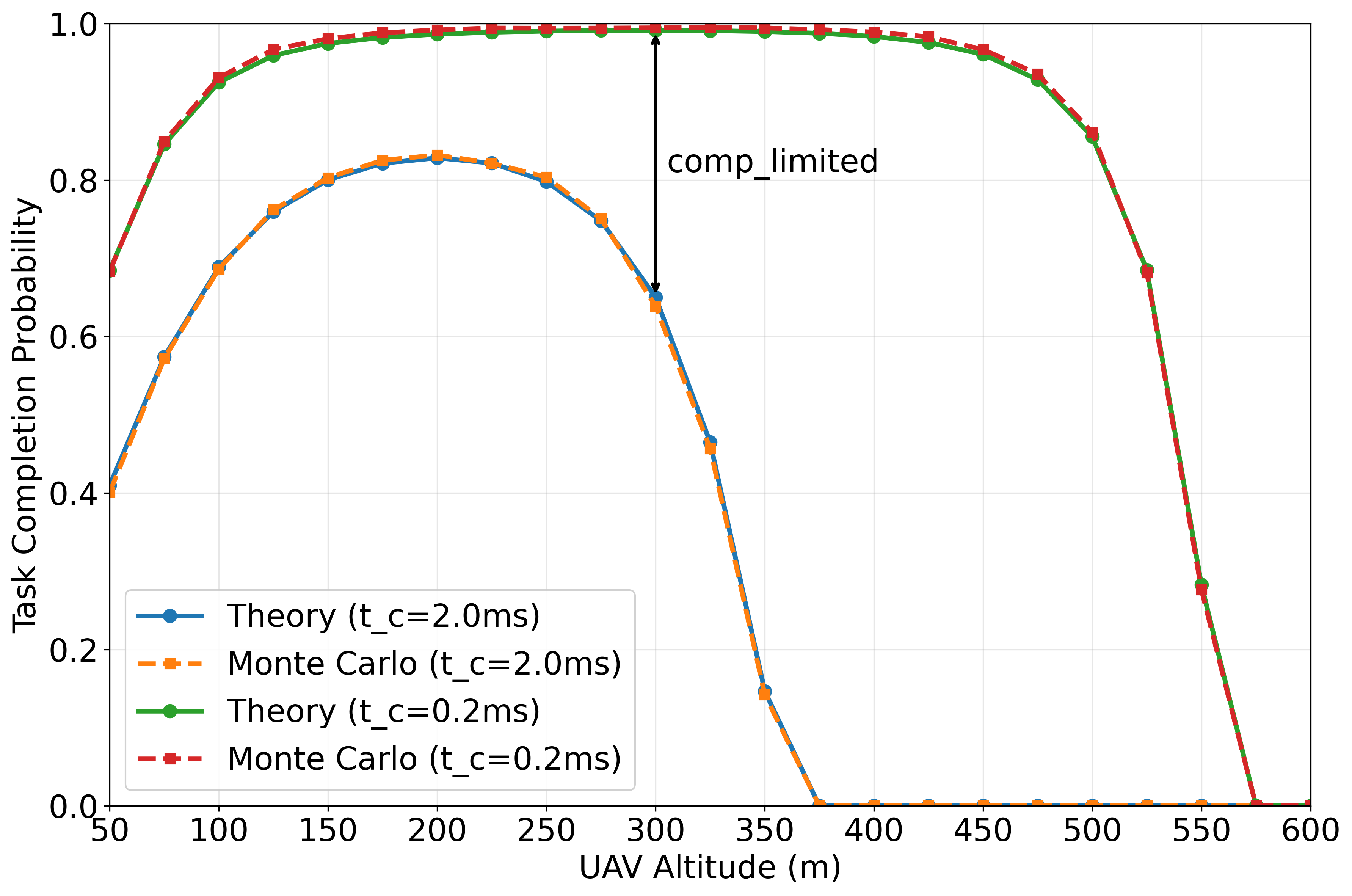}
    \caption{Task completion probability vs. UAV altitude.}
    \label{fig:success-probability-analysis}
    \vspace{-5pt} 
\end{figure}

We first validate the analytical expressions derived in Theorem~\ref{theorem:average_probability} through extensive Monte Carlo simulations (10,000 runs), each involving random sampling of 400 GUs for average task completion probability calculation. As shown in Fig.~\ref{fig:success-probability-analysis}, our theoretical analysis (\textit{Theory} curves) exhibits the close alignment with Monte Carlo simulation results (\textit{Monte Carlo} curves) across two representative computing capability scenarios: $t_c = 0.2$ ms and $t_c = 2$ ms, respectively. System performance degrades at both extremely low and high UAV altitudes.
Specifically, low altitudes result in NLoS-dominated link connections between the UAV and GUs or CNs, while high altitudes lead to excessive path loss, both of which restrict the communication coverage of the UAV and thus fundamentally limit the effective utilization of distributed computing power. Numerical results confirm the existence of an optimal UAV altitude (approximately 200m when $t_c = 2$ ms) that resolves communication bottlenecks through adaptive UAV positioning. Moreover, enhanced computing power (reducing $t_c$ from $2$ ms to $0.2$ ms) mitigates computing bottlenecks (marked as \textit{comp\_limited}), yielding altitude-dependent performance gains. The UAV-CPN architecture consequently necessitates a dynamic configuration that involves both communication and computing.

\begin{figure}[t]
    \centering
    \includegraphics[width=0.8\linewidth]{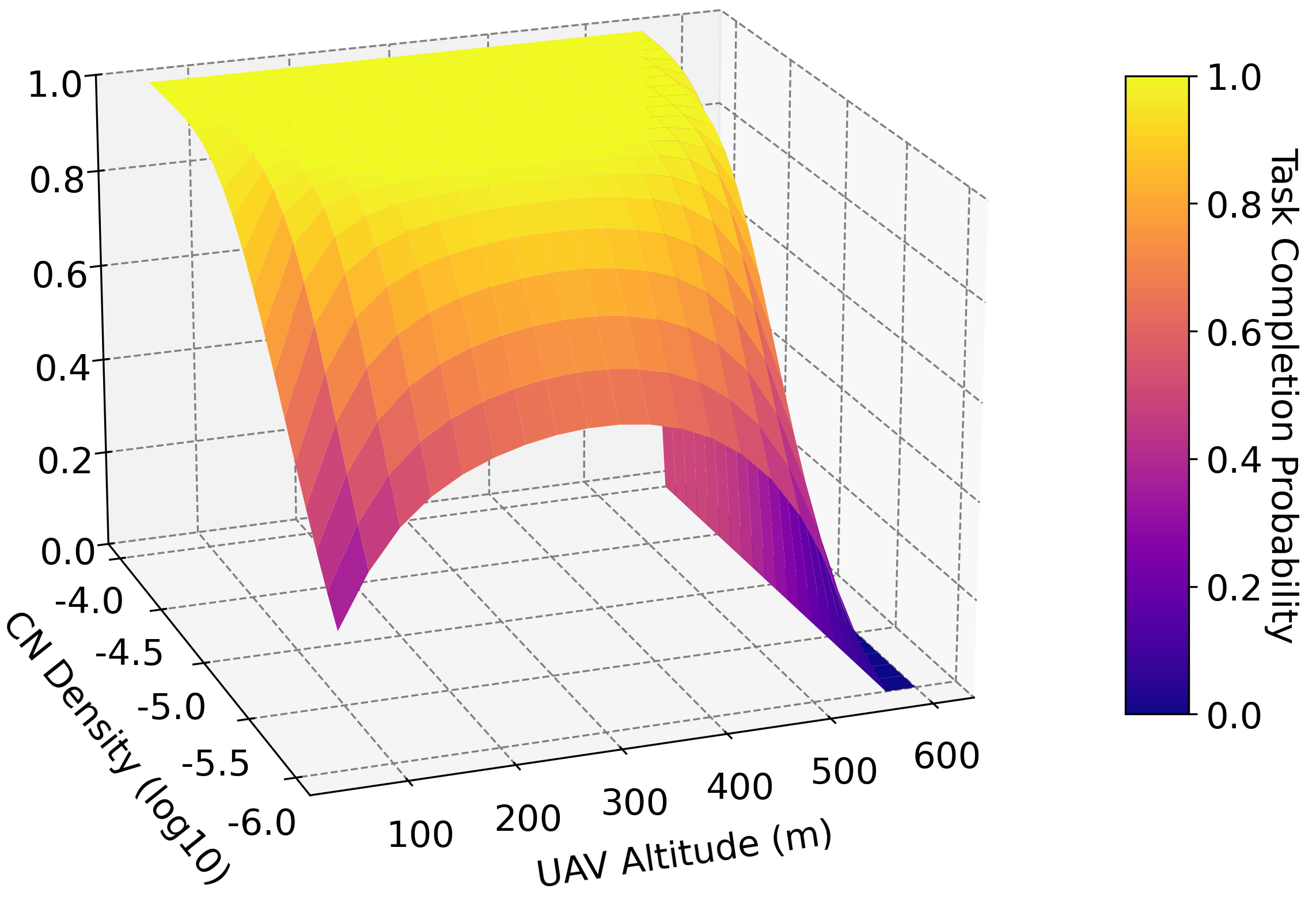}
    \caption{Task completion probability vs. CN density \& UAV altitude.}
    \label{fig:3d_altitude_density_success}
\end{figure}

We further investigate the fundamental trade-off between communication coverage and computing power by analyzing the joint effects of CN density and UAV altitude on task success probability. Figure~\ref{fig:3d_altitude_density_success} presents a 3D surface plot that reveals a nonlinear interdependency between these two critical parameters.
Specifically, the convex surface highlights different parameter sensitivities. For example, when the CN density is low, the task completion probability is highly sensitive to the altitude of the UAV. Therefore, careful design of the UAV altitude is crucial to achieve a high probability of success in the task completion under such conditions. In contrast, when the CN density is high, the altitude of the UAV can vary within a wide range while still maintaining a high probability of success in the task completion.

\begin{figure}[t]
    \vspace{-5pt} 
    \centering
    \includegraphics[width=0.8\linewidth]{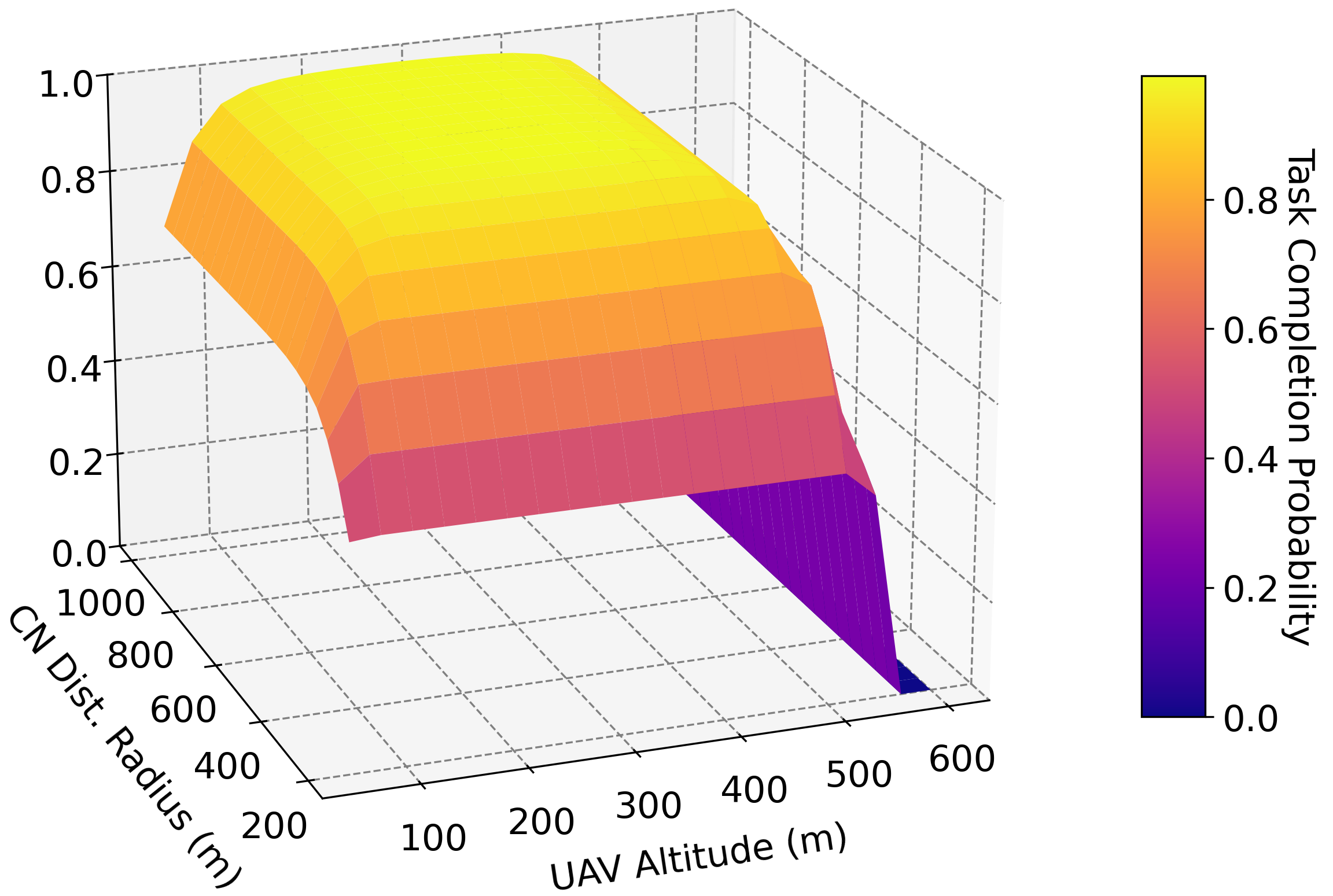}
    \caption{Task completion probability vs. CN distribution \& UAV altitude.}
    \label{fig:3d_altitude_radius_success}
    \vspace{-15pt} 
\end{figure}

Finally, we investigate the performance gain achieved through CN distribution radius expansion. Fig.~\ref{fig:3d_altitude_radius_success} presents a 3D surface plot that reveals the coupled effects of CN spatial distribution and UAV altitude on task completion probability.
For fixed UAV altitudes (e.g., at a UAV altitude of $300$m), extending CN distribution beyond the~\textit{request zone} boundaries (specifically when the distribution radius exceeds $200$m) produces substantial performance gains. Specifically, expanding the CN distribution radius from  $200$ m (corresponding to our prior work~\cite{Deng2024})  to $1000$m achieves a $2.13\times$ improvement in task completion probability ($46.65$\% $\rightarrow$ $99.14$\%), confirming our framework's capability to leverage \textit{ubiquitous computing power distribution}. For fixed CN distributions, we confirm the existence of an altitude-dependent performance maximum, demonstrating the effectiveness of~\textit{dynamic CN accessibility control} , which is consistent with our prior work~\cite{Deng2024}. While the UAV-CPN architecture introduces multi-dimensional coordination complexity, this proves to be essential for overcoming individual communication or computing bottlenecks, ultimately achieving superior performance.

\section{Conclusion}
In this paper, we have investigated the task completion probability in UAV-enabled computing power networks, where an aerial UAV relay facilitates the transmission of GUs' computing tasks to distributed CNs for real-time processing and computing. The proposed framework enables tasks generated within a constrained \textit{request zone} to be completed by CNs distributed across an unbounded geographical~\textit{service zone}, thereby increasing access to more computing power. We have derived analytical expressions to characterize the task completion probability as the main performance metric for this study. The numerical results validate the analytical expressions derived and highlight the importance of balanced resource coordination to achieve optimal system performance. By effectively managing both communication parameters (such as UAV altitude) and computing power parameters (including computing latency, CN coverage radius, and CN density), we can significantly enhance the probability of task completion. This work lays the foundation for future research on the integration of advanced computing and communication technologies into future-generation wireless networks, leading to the emerging computing power networks to be considered indispensable for AI-enabled applications.


\bibliography{deng}

\end{document}